% the second revised version 
%6.10
%corrections after 
%  last line appendix $((\b_4-\b_1)/b,\a_3-\a_1)$. 

\input harvmac.tex
\hfuzz 15pt
%\draftmode
\input amssym.def
\input amssym.tex

%%%%%%%%%%%%%%%%%%%DEFINITIONS%%%%%

% font definitions
\font\ninerm=cmr9
\font\ninei=cmmi9

\font\sixi=cmmi6

\def\({\big(}
\def\){\big)}

\def\inv{^{-1}}

 \def\frac#1#2{ {{\textstyle{#1\over#2}}}}
\def\inv{^{\raise.15ex\hbox{${\scriptscriptstyle -}$}\kern-.05em 1}}

%%%%%%%%%GREEK LETTERS%%%%%%%%%%%%%%%%
\def\a{\alpha}
\def\b{\beta}
\def\g{\gamma}
\def\d{\delta}
\def\e{\epsilon}

\def\l{\lambda}

\def\r{\rho}

\def\s{\sigma}

\def\G{\Gamma}
\def\D{\Delta}

\def\O{\Omega}

\def\o{\omega }

\def\no{\noindent}

\def\tri{\triangle}

\def\rb{ \noindent $\bullet$\ \ }

\def\tb{\tilde{\beta}}

\def\zu{\Upsilon_b}
\def\zG{\Gamma}
\def\la{\langle} \def\ra{\rangle}

\def\dC{C\kern-6.5pt I}

              \def\CC{{\cal C}}

       \def\CN{{\cal N}}

       \def\CZ{{\cal Z}}

 \def\bh{\bar h}

\input amssym.def
\input amssym.tex

%%%%%%%%%%%%%%%%% old ope

  \def\zb{\beta}

  \def\zT{\Theta}
 \def\H_{H_{1,2}}  \def\zT_{\Theta_{1,2}}
 \def\O_{O_{1,2}} \def\bH_{{\bar H}_{1,2}}
 
 \def\V_{V_{1,2}}
 \def\D_{D_{1,2}} \def\bD_{{\bar D}_{1,2}}

% caligraphic:

       \def\cD{{\cal D}}

     \def\cD_{{\cal D}_{1,2}}
 \def\bcD_{{\bar {\cal D}}_{1,2}}

% doubled letters:

   \def\dC{I\!\!\!\!C}

% convenient macros:

% local macros:

%\def\hl{\hat{\lambda}}

\def\dal{
\vbox{
\halign to5pt{\strut##&
\hfil ## \hfil \cr
&$\kern -0.5pt
\sqcap$ \cr
\noalign{\kern -5pt
\hrule}
}}\ }

%%%%%%%%%%%%%%%

%%%%%%%%%%%%

\def\rb{$\bullet\,$}

\def\tF{\tilde F}
%%%%%%%%%% References %%%%%%%%%%  

\lref\FLuk{ V.A. Fateev and S. Lukyanov, The models of two-dimensional conformal quantum field  theory with $Z_n$ symmetry, Int. J. Mod. Phys. {\bf A3} (1988) 507.}
\lref\FBas{
V.A. Fateev and  P. Baseilhac, 
Expectation values of local fields for a two-parameter family of integrable models and related perturbed conformal field theories, Nucl. Phys. B532 (1998) 567-587, 	arXiv:hep-th/9906010.}
\lref\FLc{ V.A. Fateev and A.V. Litvinov, Correlation functions in conformal Toda field theory I,
 JHEP {\bf 11} (2007) 002, arXiv.0709.3806.}
\lref\FLd{ 
V.A. Fateev and A.V. Litvinov, Correlation functions in conformal Toda field theory II,
JHEP {\bf 09} (2009) 033, arXiv:0810.3020.}
\lref\FLa{ 
V.A. Fateev and A.V. Litvinov, Multipoint correlation functions in Liouville field theory and minimal Liouville gravity,
Theor.Math.Phys.{\bf 154} (2008) 454-472, arXiv:0707.1664.}
\lref\MP{
V.  Mitev and   E. Pomoni,  Toda 3-point functions from topological strings, JHEP {\bf 06} (2015) 049,  	arXiv:1409.6313; \
M.  Isachenkov, V.  Mitev,  E. Pomoni,  Toda 3-point functions from topological strings II,	arXiv:1412.3395. }
\lref\AGT{ L.F. Alday, D. Gaiotto and  Yu. Tachikawa,  Liouville correlation functions from four-dimensional gauge theories,  
{\it Lett. Math. Phys.} {\bf 91} (2010) 167, arXiv:0906.3219.}
\lref\Wyl{N. Wyllard, A(N-1) conformal Toda field theory correlation functions from conformal $N=2$
SU(N) quiver gauge theories, JHEP 0911(2009) 002, arXiv:0907. 2189.}
\lref\MS{G. Moore and  N. Seiberg, Classical and quantum conformal field theory, Comm.  Math. Phys. {\bf 123} (1989) 177-254.}
\lref\JMOb{M. Jimbo, T. Miwa, M. Okado, Solvable lattice models related to  the vector representation of  classical simple Lie algebras,  Comm. Math. Phys. {\bf 116} (1988) 507-525.}
\lref\JMO{M. Jimbo, T. Miwa and M. Okado,
Solvable lattice models whose states are dominant integral weights of $A_{n-1}^{(1)}$,
 Lett. Matt. Phys. 14 (1987) 123-131.}
 \lref\JW{R.  Janik and  A. Wereszczynski, Correlation functions of three heavy operators - the AdS$_5$ contribution, JHEP {\bf 12} (2011) 095, arXiv:1109.6262.}
  \lref\KK{Y. Kazama and S. Komatsu, Three point functions  in the SU(2) sector at strong coupling,
 JHEP {\bf 03} (2014) 052, arXiv:1312.3727.}
 \lref\FGPP{ P. Furlan, A.Ch. Ganchev, R. Paunov and V.B. Petkova,
% Reduction of the rational spin $sl(2,C\!\!\!\!I)$ WZNW conformal
%theory,
% Phys. Lett. {\bf B267}  (1991) 63-70;  \qquad \qquad  \qquad  \qquad  
%P. Furlan, A.Ch.  Ganchev, R. Paunov and V.B. Petkova,
 On the Drinfeld-Sokolov reduction of the Knizhnik-Zamolodchikov equation,
in the   Proceedings of  the {\it Workshop on Low Dimensional Topology and
Quantum Field Theory},  Newton Institute, Cambridge, Sept., 1992, 
edt. H. Osborn, NATO ASI Series B, v. {\bf 315} p. 131-141 (Plenum Press,
New York, 1993).  }
 \lref\AlZ{Al.B. Zamolodchikov, The three-point function in the minimal
Liouville gravity, Theor. Math. Phys.  {\bf 142} (2005) 183-196,
%; On the three-point function in minimal Liouville gravity,
hep-th/0505063.}
 \lref\KPa{
  I.K. Kostov and V.B. Petkova, Bulk correlation
functions in 2D quantum gravity,
Theor. Math. Phys. {\bf 146} (1) (2006), 108-118, hep-th/0505078.}
 \lref\LMRS{ S. Lee, S. Minwalla, M. Rangamani and N. Seiberg, Three point functions of chiral operators in $D=4, \CN=4$ SYM at large $N$, Adv.Theor. Math. Phys. 2 (1998) 697-718,
 hep-th/9806074.}
\lref\Th{ C. Thorn, Liouville perturbation theory, Phys. Rev. {\bf D 66}, 027702 (2002), hep-th/0204142.}
\lref\FMMR{D.Z.  Freedman, S.D.  Mathur, A. Matusis,  L. Rastelli,
Correlation functions in the CFT$_d$/AdS$_{d+1}$ correspondence,
Nucl. Phys. 546 (1999)   96-118, hep-th/9804058.}
\lref\AF{G.Arutyunov, S.Frolov, Some cubic couplings in type IIB supergravity on Ad$S_5 \times S^5$  and three-point functions in SYM$_4$ at large N, 	Phys.Rev. D61 (2000) 064009, hep-th/9907085.}
\lref\DPak{A. Dabholkar and A.  Pakman, Exact chiral ring of AdS$_3$/CFT$_2$,  Adv.Theor.Math.Phys.{\bf 13} (2009) 409,
arXiv:hep-th/0703022. }
\lref\GabK{M.R. Gaberdiel and I. Kirsch, World sheet correlators in Ad$S_3/$CFT$_2$, JHEP 04 (2007) 050, hep-th/0703001.}
\lref\ASY{O. Aharony, J. Sonnenschein and  S. Yankielowicz, G/G models and $W_N$ strings, Phys. Lett. {\bf B 289} (1992) 309-316, hep-th/9206063.} 
\lref\ZZ{A. Zamolodchikov and Al. Zamolodchikov, Structure constants and conformal bootstrap in Liouville field theory, Nucl.Phys. {\bf B 477}   (1996) 577-605, hep-th/9506136.}
\lref\BFPS{ P. Bozhilov, P. Furlan, V.B. Petkova and M. Stanishkov, 
 On the semi-classical 3-point function in AdS$_3$, 
 Phys. Rev. {\bf D 86} (2012)  066005,  arXiv:1204.1322 [hep-th]. }
 \lref\BPT{Z. Bajnok, L. Palla, G. Takacs, $A_2$ Toda theory in reduced WZNW framework and the representations of $W$ algebra, Nucl. Phys. {\bf B385} (1992) 329-360, hep-th/9206075.}  
 \lref\Baj{Z. Bajnok, New approach to the correlation functions  of $W$-algebras, hep-th/9906185. }
 \lref\Wat{G.M.T. Watts, Fusion in the $W_3$ algebra, Comm. Math. Phys.  {\bf 171} (1995) 87-98,  hep-th/9403163.}
 \lref\BW{P. Bowcock and G.M.T. Watts, Null vectors, 3-point and 4-point functions in conformal field theory, Theor. Math. Phys. {\bf  98}  (1994) 350-356, 
%Teor.Mat.Fiz. 98 (1994) 500-508
hep-th/9309146.}
\lref\FKW{E. Frenkel, V. Kac and   M. Wakimoto,  Characters and fusion rules for W-algebras via quantized Drinfeld-Sokolov reduction, Comm. Math. Phys. {\bf 147} (1992)  295-328. }

  %%%%%%%%%
 
 \overfullrule=0pt

\Title{\vbox{\baselineskip12pt\hbox
{}\hbox{}}}
{\vbox{\centerline
 {On some  3-point functions  in the  $W_4$ CFT}
 \medskip
 \centerline{ 
 and related braiding matrix } 
}}
 \centerline{  P. Furlan$^{*}$ and V.B. Petkova$^{**}$  }
\vskip 5pt
\medskip
\bigskip

 \centerline{ \vbox{\baselineskip12pt\hbox
{\it ${^{*)}}$Dipartimento  di Fisica 
 dell'Universit\`{a} di Trieste, Italy;}
 }}
  \centerline{ \vbox{\baselineskip12pt\hbox
{\it ${^{**)}}$Institute for Nuclear Research and Nuclear Energy, }
}}
 \centerline{ \vbox{\baselineskip12pt\hbox
 {\it Bulgarian Academy of Sciences, Sofia, Bulgaria}
 }}

 \bigskip
 \bigskip

 %%%%%%%%%%%%%%%%%%%%%%%%%%%%%%%%%%%

\noindent

\no 
We construct  a
 class of   3-point  constants    in the $sl(4)$ Toda conformal theory  $W_4$, 
extending the 
examples in Fateev and Litvinov \FLc.  
Their knowledge
allows to determine the  
braiding/fusing  matrix 
transforming  4-point conformal blocks  
of one fundamental, 
labelled by  the 6-dimensional 
 $sl(4)$ representation, and three  partially degenerate 
  vertex operators.
It is a 
 $3 \times  3$ submatrix  of the generic $6 \times 6$ fusing matrix consistent with the fusion rules for  the particular class of  
 representations.  
 We check a 
   braiding relation which   
 has wider applications 
to   conformal models   
with   $sl(4)$ symmetry.
The 3-point  constants in  dual 
 regions  of central charge are compared 
 in preparation for a BPS like  relation in the 
$\hat{sl}(4)$ WZW model.

%%%%%%%%%%%%%%%%%%%%%%%%%%%%%%%%%%%

\bigskip
\bigskip\bigskip
\bigskip

\no
-------------------------------------------------------------
\medskip
\no
furlan@ts.infn.it, petkova@inrne.bas.bg
\Date{}

%%%%%%%%%%%

\newsec{Introduction}

The 2d conformal field theories (CFT) related to the $sl(2)$ algebra, like the Virasoro,  the WZW models with the affine sl(2) KM  algebra and their supersymmetric extensions,  are by now well established. This includes explicit expressions for 
  basic data as the  operator product expansion  (OPE) coefficients (3-point functions) and  the braiding/fusing matrices transforming conformal blocks.
Much 
  less  
is  known about  these structures in the CFT  with  higher rank symmetries,  
 although a  considerable progress in Toda CFT \FLuk\ was made 
 by Fateev and Litvinov (FL) \FLc,  \FLd. 
Further advances in the field  are important  for the development of the higher rank 2d CFT
as well as for   potential applications in the string theory side of the AdS/CFT correspondence.

 In the  free field (Coulomb gas) approach the OPE constants are represented by 
  complicated  integrals   which have  to be computed explicitly before analytic continuation. The alternative derivation of functional relations   arising from locality (crossing symmetry) of  particular  4-point functions  involving  degenerate  vertex operators 
 requires  the knowledge of  fundamental braiding/fusing matrix elements, 
 which      in general are also part of the problem.
 
In \FLc,  \FLd\   
Fateev and Litvinov  developed a general method of recursively computing certain class of conformal integrals and gave explicit  examples of 3-point constants.\foot{Apart from  these  traditional 2d methods a  novel  approach to the computation of the 3-point constants  is   provided by the 
  (5d version of the) AGT-W relation \AGT, \Wyl,  see   \MP,  
where the main example  of \FLc\ has been recently   reproduced, as well as  references therein.}
In  the case of  Toda $W_3$ theory  they   have as well computed the fundamental fusing matrix directly from the integral representations of the 4-point blocks; some partial results  in the general $W_n$  case  were also obtained.

In this paper we are dealing with the sl(4) Toda conformal theory $W_4$.
The 3-point  functions  known so far  involve one 
 vertex operator  $V_{\b} $ with  a degenerate  charge $\b$
proportional to the fundamental weight 
$\o_1$, or $\o_3=\o_1^*$, i.e., the highest weight of the  4-dimensional  $sl(4)$  representation. 
Our focus instead is on the symmetric representations $ \b=\b^*$
 and, in  particular,
 $\b=- k  \o_2 b$, where $\o_2$ is the highest weight of the 6-dimensional fundamental $sl(4)$ 
representation and $k$ is arbitrary\foot{These highest weights correspond to scalars in 
 the context of  4d  conformal group  
  representations, where in general  the (nonnegative, integer) components $2j_i=-(\b,\a_i)/b, i=1,3$
 label the $SL(2,C)$ spins,  while $\tri=(\b, \o_2)/b  $ corresponds to the 4d conformal dimension.}.
 The real parameter  $b$ parametrises   Toda central charge.
In section 2 we present   a  3-point OPE constant  for two partially degenerate ("4d scalars")  and one symmetric representations 
\eqn\clasreps{
(\b_1, \a_i)=0= (\b_2, \a_i)\,, \ {\rm for} \ i=1,3  \,, \ (\b_3,\a_1)=(\b_3,\a_3)\,.
 }
The 3-point constant is obtained 
by  deriving and solving a recurrence relation for the  
corresponding Coulomb gas integrals along the method  of \FBas, \FLc, which is 
 then analytically continued.  

In section 3 we  use this data to derive   the 
  fusing matrix  $F$ transforming the corresponding 4-point conformal blocks with one fundamental vertex $V_{-\o_2 b}$.
  Here we follow a path somewhat opposite to the  standard consideration in which  - given the  fusing matrix,  one solves  for the 3-point constants the system of 
equations implied by  locality of the  4-point function. We shall not need the explicit integral realisation of this particular Toda 4-point function with three more 
 partially degenerate  representations of the  type   $\b_a=- k_a  \o_2 b\,, a=1,2,3\,.$
 In the intermediate channels  appear also vertex operators with symmetric weights so that in the equations the more general  constants of the type  \clasreps\ derived in section 2 are needed. 
 The restriction to chiral vertex operators  $V_{\b_a}$ of such particular highest weights effectively restricts the braiding/fusing matrix  to a $3 \times 3$  submatrix; its  matrix elements 
are explicitly described. 

Finally in this section  we
 check a braiding identity,  which is equivalent to a standard identity for the modular group on the sphere with 4 holes. This relation 
imposes  restrictions  solely on certain  products of $F$ matrix elements and 
allows  
 in principle for more general solutions for  the  individual $F$ matrix elements than the ones computed in the $W_4$ CFT. The semi-classical "heavy charges" limit of the  identity 
 is a particular $sl(4)$ analog of the one exploited in the strong coupling  $sl(2)$   sigma model constructions 
 in \JW, \KK. 
 This suggests that the explicit expressions for the products of the fusing matrix elements
extracted   from 
Toda CFT  (or their closely related WZW model counterparts) may eventually be used as a first step in  higher rank generalisations of that  work.

In the last section 4  we compare  the 3-point constants  in two regions of the central charge, $W_4$ analogs of the  two Virasoro theory  ingredients of Liouville gravity (non-critical string theory)   with $c>25$ (Liouville) and $c<1$ ("matter"). 
 We discuss  a BPS-like relation 
for the two sets of weights which is intrinsic for the vertex operators of the $\hat{sl}(4)$ WZW models, related to Toda theory by the quantum Hamiltonian
Drinfeld Sokolov (DS)  reduction. Although it seems that there are  no direct $W_4$ analogs  of the physical fields of Liouville gravity, we
 show that     the product of the two 3-point Toda constants with weights subject to the BPS constraint trivialises 
in the  semi-classical "light" charges" limit. 
Furthermore  we speculate on the possible implications for the  related WZW 3-point correlators 
(the determination of which is still an open problem)  and make a comparison 
 with computations  of related 3-point  correlators  in the supergravity approximations of the $AdS_5 \times S^5$ strings
 \FMMR, \LMRS.

  The Appendix contains some details of the computation of the  3-point functions, including 
one  slightly more general  constant  not presented in section 2, as well as 
an alternative Coulomb like  representation of the 4-point functions discussed in section 3. It reveals a connection to certain Liouville correlators.

\newsec{3-point $W_4$ constants}

We consider the $W_4$ CFT with central charge 
\eqn\tod{
c_T= 3 (1+20 Q^2)=3\(41+ 20(b^2+{1\over b^2})\) > 243\,, 
\ Q={1\over b}+b\,,
}
for real values of the parameter $b$. We shall skip the detailed presentation of the basics of Toda conformal theory and 
the  free 
 field (Coulomb gas) representation  of the  correlation functions: the reader is referred to  \FLc, as well as to the original paper  of Fateev and Lukyanov \FLuk, formulated in the   
 dual region of central charge with $b \to i b$ in \tod.

 The OPE constant of 2d scalar vertex operators is
$$
c(\b_1,\b_2\,,  2\r Q- \b_3) =\lim_{x_3\to \infty}(x_3^2)^{2\tri(\b_3)}\,
\la  V_{2\r Q-\b_3}(x_3) V_{\b_2}(1) V_{\b_1}(0) \ra_{\rm Coulomb}
$$
where the conformal dimension is given by the $sl(4)$ inner product
 \def\bdot{\star}
 \eqn\dimqr{
\tri(\b)= {1\over 2}(\b, 2\r Q-\b)\,   \  
 }
and $ \ \r =\sum_{i=1}^3 \o_i$ is the  Weyl vector. The dimension \dimqr, as well as the two other  $W_4$ quantum numbers,  are invariant with respect
to  an action of the Weyl  reflection group  
\eqn\weyltod{
w\bdot \b= 
Q \r +w(\b-Q\r )\, 
}
so that any of the vertex operators $V_{w\bdot \b}$ represents  the same field.
 The Coulomb gas representation of the OPE constant is defined for the
charge conservation condition 
\eqn\cconsc{
\b_{12}^3+b\sum_i s_i \a_i :=\b_1+\b_2-\b_3+b\sum_i s_i \a_i=0 \  \Rightarrow  \ b s_i=-(\b_{12}^3,\o_i)\,.
}
The integers $s_i$ in front of the simple roots $\a_i$ in \cconsc\ count the number of screening charge  vertex operators $V_{\a_i b}  (z,\bar z)\,, $   
 $ i=1,2,3$,  from  the interaction term of Toda action.
 These operators are spinless fields of dimension $\tri(\a_i b)=1$. Formula 
 \cconsc\ 
describes a  generic $sl(4)$  type fusion rule in which $\b_3$ is obtained by a shift of, say,  $\b_1$ with the weight diagram $\Gamma_{-\b_2/b}=\{-\b_2/b - s_i\a_i\}$ of the representation of highest weight  $-\b_2/b$,  times $(-b)$.  
 
 The OPE constant is given by a $\sum_i s_i -$ multiple 2d integral  $I_{s_1,s_2,s_3}(\b_1,\b_2)$
 (formula (1.33) of \FLc\ recalled in (A.1) below).
  We  compute this  integral in the particular case  when the three highest weights are chosen as in
  \clasreps.
 The components $(\b_a, \a_2)$ of the  weights in \clasreps\ take arbitrary values, subject of the condition  \cconsc; the latter implies  that $s_1=s_3$ and we shall assume that  the integer $l:=-(\b_3, \a_1)/b=s_2-2s_1$ is  nonnegative.  

  We shall skip the detailed computation of the OPE constant since it follows  straightforwardly 
 the steps of the method explained in \FLc, 
 which is based on the use of a 
  $sl(2)$ type duality formula \FBas\  in order   to derive recursion  relations for  $sl(n)$ Toda multiple 2d integrals; see the Appendix for  a short summary of the procedure.
    In our case after $s$ steps one gets  an integral
of type $$I_{s_1-s,s_2- 2 s,s_3-s}(\b_1+s \o_2 b,\b_2+s\o_2 b)$$ so that setting $s=s_3=s_1$ the integral is reduced to known  Liouville Coulomb  integral.  
In particular for $\b_2=-\o_2 b$ the resulting formula reproduces the structure  constants  $c_h:=c(\b_1, -\o_2 b, 2\r Q- (\b_1- h b))$ of the fusion of $V_{ \b_1}$ with the fundamental field  $ V_{-\o_2 b}$ corresponding to  three  of the six  points of  the weight diagram $\Gamma_{\o_2}$, i.e., 
\eqn\weidi{\eqalign{
&\b_3=\b_1 - h b \   {\rm  with } \ \cr 
& h= \o_2\,; \ 
h=
\o_2-\a_2=w_2(\o_2)\,;\ 
h=-\o_2 =\o_2- 2\a_2-\a_{1}-\a_3
=w_{2312}(\o_2)\,. 
}}
These are the three weights of $\Gamma_{\o_2}$ preserving the symmetric type $\b_3=\b_3^*$ with $l\ge 0$. 
The expressions for these OPE constants reproduce special cases 
of the general formula (1.51)    of  \FLc\    
valid    for  arbitrary  $\b_1$.  For  
the  partially degenerate   weights of type $(\b_1, \a_i)=0\,, i=1,3$
 the remaining three OPE constants in \FLc\  vanish,  in agreement with the vanishing of the corresponding
 $sl(4)$  tensor product decomposition  multiplicities:  the $sl(4)$ (or  $\hat{sl}(4)$) Verma modules of  
 highest weights $\l=-\b/b$  with non-negative  integer components $l^{(i)}= (\l, \a_i), i=1,3$ have two singular vectors,
 whose factorisation imposes additional restrictions on the fundamental  fusion rule. 
In particular in  the case $(\l_1,\a_1)=0=(\l_1,\a_3)$  for 
each of the three  weights   $h= \pm (-\o_1+\o_3)\,,   \o_2- (\o_1+\o_3)\in \Gamma_{\o_2}$ there is an odd  Weyl  group element  $w_1$, or $ w_3$, or both,   the shifted action of which 
   keeps 
   $\l_3=
 \l_1+h$  invariant, 
   a property which does not depend on the value of $(\l_1, \a_2)$,  and    which implies the vanishing of the corresponding fusion multiplicities.\foot{  The  sl(4) pattern of the $W_4$  
 fusion rules multiplicities   is independently proved in the particular case of  integer  dominant weights $-\b/b$  extrapolating   the rational $b^2$ result in \FKW, derived by reduction of the $\hat{sl}(n)$  WZW Verlinde formula.
The fusion rule of $f=-\o_2 b$  with representations of generic highest
weight  can  be derived  algebraically  accounting for the  factorisation of the null states in the corresponding  completely degenerate $W_4$ Verma module.
The three independent singular vectors  
- two at level $1$ and one at level $2$,  inherited via the  quantum DS reduction from  the  
 singular vectors of $\hat{sl}(4)$ module,  %corresponding to the simple roots,  
 are not sufficient.
  Together with  the two projective Ward identities corresponding to the zero modes $W_0^{(j)}$ of the   spin $j=3$ and $j= 4$ currents, they provide 
five relations which  eliminate the 3-point matrix elements containing  the  
negative  modes $W_{-k}^{(j)},  1\le k \le  j-1$: the latter determine the action on the fields of all  higher  negative modes 
 $W^{(j)}_{-n}, n\ge j $. To derive the fusion rule itself
one needs to explore  the factorisation of  three more descendent null states, presumably at levels up to $5$, as suggested  by the classical KZ equation for this representation;  see  the general discussion in  \BPT,     \Wat,  \FLc\  applied to  $W_3$ examples.
}

\bigskip 
\noindent
$\bullet$  Our  next step is the standard analytic continuation of the OPE constant, to be denoted $C(\b_1, \b_2, 2\r Q- \b_3)$ for weights of the type \clasreps\  not restricted by \cconsc,  
so that  the Coulomb gas OPE constant is reproduced as a double  residue
\eqn\resdoubl{ c(\b_1, \b_2, 2\r Q- \b_3)=
 {\rm res}_{ (\b_{12}^3, \o_2-2\o_1)=-(s_2-2s_1) b} \, {\rm res}_{(\b_{12}^3,\o_1)=-s_1b}C(\b_1, \b_2, 2\r Q- \b_3)
}
where $s_1$ and $s_2-2s_1$ are nonnegative integers. 

We shall write down the  related   formula  for  
$\b_3 =\b^*_3$ replaced by $  2\r Q-\b_3$, equivalently obtained by multiplication with the reflection amplitude $R(\b_3)$   corresponding to the longest Weyl group element  $w_{121321}$  \FLc\ 
\eqn\reflampl{
R(\b)=(b^{2(1-b^2) }\l_T)^{(2\r Q-2\b, \r )\over b} \prod_{\a>0}{\zu((\r Q-\b, \a))\over \zu((\b-\r Q, \a)) }\,,
}
namely,
  \eqna\formT$$\eqalignno{
& C(\b_1, \b_2, \b_3)= R(\b_3)\,  C(\b_1, \b_2, 2\r Q- \b_3)\cr
&= \(b^{2(1-b^2)}\l_T\)^{ {(2\r Q- \b_{123},\r)\over b}}\,  
{\prod_{\a=\a_1,\a_{14}}\zu((\b_3-\r Q,\a)+Q)\over \prod_{\a=\a_1,\a_{13}}\zu((\b_3-\r Q,\a))}\times \cr
&\prod_{a=1,2}{\zu((\b_a,\a_2))
\over  \zu((\b_{123}-2\b_a,\o_2-\o_1)) }{\zu(b)\, \zu((\b_3,\a_2))   \over  \zu((\b_{12}^3,\o_1)) \zu((\b_{123},\o_2-\o_1) -2Q) }\times  &\formT{}
\cr
&\prod_{a=1,2}{  \zu((\b_a-\r Q,\a_{24})+Q)  \over   
\zu((\b_{123} -2\b_a,\o_1)-Q)} {\zu(b)\, \zu((\b_3-\r Q,\a_{24})+Q) \over  
 \zu((\b_{12}^3,\o_2-\o_1)-Q) \zu((\b_{123},\o_1)-3Q) }\,.\cr
 &
 }$$
 Here
$\l_T$ is proportional  to the Toda cosmological  constant,  $\l_T=\pi \mu_T \g(b^2) $. 
Recall that 
 $\zu(x)$ is an entire function with zeros at $x=-nb-m/b\, $ and $x=Q+nb+m/b$, $ n,m\in \CZ_{\ge 0}\,, $ satisfying  
the functional  relations
\eqn\fnct{
\zu(x+b^{\epsilon})=  \g(xb^{\epsilon})\,b^{\e(1-2x b^{\e})} \,\zu(x)\,, \epsilon=\pm 1\,, 
}
$\g(x)=\G(x)/ \G(1-x)$. In the products over positive roots  in \formT{} the root  $\a_1$
can be replaced  with $\a_3$  (and $\a_{13}=\a_1+\a_2$  - with $\a_{24}=\a_2+\a_3 $)
 since $\b_3=\b_3^*\,, \r=\r^*$. The ratio in the second line 
 of \formT{} produces  a  finite constant for $(\b_3,\a_1) \to -lb\,, l - $nonnegative integer so that \formT{} has sense for such values of $\b_3$ whenever  the components $(\b_a, \a_2)$ of the three weights are generic.
 We have also used that  the two weights $\b_1,\b_2$ have zero components   $(\b_a,\a_i)=0$ for $i=1,3$ in order to write \formT{}  in a form which makes it explicitly symmetric  when  the third weight $\b_3$ is also chosen of this type, i.e., $l=0\,,$ as we shall need it below: in that particular case 
 $(\b,\o_2-\o_1)=(\b,\o_1)$ in  all 
  products of this type in the denominators in the last two lines. 
 
\medskip
Using \weyltod\   the terms in  \formT{} 
depending on the three vertices, i.e., the eight  $\zu$-factors in the denominators of the last two lines,  can be also written as points on an  orbit
of the Weyl group acting on the three weights (as  discussed, e.g.,    in \MP\ for  the FL example) 
\eqn\weylrep{
{\Big( \zu((\b_{123}-2\r Q,  \o_1) \zu((w^{(3)}_{13}\bdot\b_{123} -2\r Q,\o_1)\,
 \zu((w^{(3)}_{121321}\bdot \b_{123}-2\r Q,\o_1)) \Big)^{-1}
\over 
\prod_{j=1}^3  \zu((w^{(j)}_{2132}\bdot \b_{123}-2\r Q,\o_1))\,    
\prod_{i=1,2} \zu((w^{(i)}_{2132}w_{13}^{(3)}\bdot \b_{123}-2\r Q,\o_1))}
}
 
\medskip

\noindent
where  
$w'^{(1)}w^{(3)}\bdot \b_{123}=w'\bdot\b_1+\b_2 + w\bdot \b_3\,,$ etc..

\newsec{ Locality, fusing matrix, braiding identity} 
Consider the local 4-point function $\la V_{f} V_{\b_1}V_{\b_2} V_{\b_3} \ra\, $ of  primary spinless operators $V_{\b}(z,\bar z)$ one of which is labelled by  a fundamental highest weight, 
in our case $f=-b\o_2$. The most interesting for the applications are the cases in which the remaining  three  primary fields have 
 highest weights $\b_a$ 
 with  nonnegative integer components $l_a^{(i)}= -(\b_a, \a_i)/b\,, i=1,3$ and generic  $(\b_a,\a_2)$
(or,  any of their  Weyl group related  values providing equivalent vertex representations). 
These representations arise from doubly reducible  Verma modules with two singular vectors. The  projective Ward identities and the factorisation of all    singular vectors - as well as the  descendent null states in the fundamental representation $f$
   give restrictions on the conformal blocks reducing the space of descendent states described  in terms of powers of the  modes $W_{-k}^{(j)},  1\le k \le  j-1$, to a finite dimensional subspace, see 
  \BW, \Baj, \FLc\  for different approaches to this problem.
  Instead of the detailed analysis of  this space one can  give, as in  \FLd,  an alternative argument showing that 
  at least a subclass of 
 these  4-point functions
 admit  an  integral  Coulomb gas like  representation, see the Appendix. 
 This indicates that all descendent  states  of the above space are eliminated so that the fusion channels of these highly degenerated 4-point functions follow the $sl(4)$ pattern dictated by the completely degenerate field   $V_{-\o_2 b}$; 
 in what follows we restrict to this subclass of 4-point functions. 
  In the case when all $l_a^{(i)}=0\,, i=1,3\,, a=1,2,3$  - which is the main case under consideration below,  the  alternative Coulomb representation allows to identify 
  the linear  differential equation satisfied by the 4-point function. 

 \medskip
 
 \noindent
 $\bullet$ 
The 4-point function admits different  equivalent  diagonal  decompositions in   conformal blocks. They are    related by   braiding transformations, 
i.e., matrix  realisation of the braiding group with generators
$e_i\,, i=1,2,3$ on the plane (Riemann sphere)  with 4 holes;  
$e_i $ is exchanging the chiral vertex  
 operators at the $i$-th and $i+1$-th points  and 
the   notation refers  to the fixed  ordered points, not to the labels of 
 the concrete interchanged operators.  In particular the generators  $e_2$
 (for the above order of the corresponding chiral vertex operators) is represented by non-trivial braiding matrix $B$ proportional to the fusing matrix $F $
\eqna\brai$$\eqalignno{
& B_{\beta_1\!-\!h_s b,
\beta_2\!-\!h_t b}\left[\matrix{\beta_1&\beta_2\cr  f &\b_3} \right](\epsilon)\!=\!
e^{i\pi  \epsilon( \tri(\b_3)\!+\!\tri(f)\!-\!\tri(\beta_1\!-\!h_s b)\!-\!\tri(\beta_2\!-\!h_t b))}
 F_{\beta_1\!-\!h_s b,
\beta_2\!-\!h_t b}\left[\matrix{\beta_1&\beta_3\cr   f &\b_2} \right] \,, \cr
&& \brai{} \cr
& B_{\g,\d}\left[\matrix{\beta_1&\beta_2\cr  \b_4&\b_3} \right](\epsilon)
 B_{\d,\g'}\left[\matrix{\beta_2&\beta_1\cr  \b_4&\b_3} \right](-\epsilon)=\d_{\g \g'}
 }$$
while  $e_1$ and $e_3\,,$ which exchange  the operators in the first two, respectively last two,  fixed points,  
reduce due to triviality of $F$, to  diagonal matrices. 
Locality (symmetry under exchange of two 2d  fields)  requires that the function is invariant under such transformations relating different diagonal chiral decompositions.    This results in equations involving fusing matrix elements and products of 3-point constants. In the case under  consideration  
the equations  take  the form of a  finite sum.
% according to the $sl(4)$ tensor product  of the representation $f$.
E.g.,  for the exchange of $V_{\b_1}$ and $V_{\b_2}$ they read
  \eqn\loceq{\eqalign{
&\sum_{h_s\in \G_{\o_2}}
 {c_{h_s}(\b_1)  C(\b_1 -h_s b, \b_2,\b_3)\over
 c_{h_t}(\b_2)\, C(\b_1, \b_2-h_t b, \b_3)}
 F_{\b_1 -h_s b, \b_2 -h_t b}\,
 F_{\b_1 -h_s b, \b_2 -h_u b} =\delta_{h_t, h_u}\cr
 &= \sum_{h_s\in \G_{\o_2}} (F^{-1})_{h_t\, h_s} F_{h_s\, h_u}\,.
 }}
Here $c_{h_s}(\b_1)$ 
 is a shorthand notation for the OPE constant 
 $c(-\o_2 b, \b_1, 2\r Q- (\b_1-h_s b))$, see the  general formula (1.51) in  \FLc.
 In particular $c_{h=\o_2}=1$. In general $h_s$ stands for the weights of the weight diagram $\Gamma_{\o_2}$ of the 6 dimensional  representation, but for our restricted set of highest weights $\b_a=-k_a \o_2 b\,, \, a=1,2,3$, three of the OPE coefficients $c_{h_s}$ given in \FLc\ vanish,   as discussed above, 
 so we are left with summation over 3 of the weights, as given in \weidi.  
 A  shorthand notation for the matrix 
 $F_{h_s, h_t}= F_{\b_1 -h_s b, \b_2 -h_t b}$ in the last equality in \loceq\ is used.

 As indicated in the r.h.s. of \loceq\
  the matrix formed by the ratio of constants times $F$ can be identified with the  inverse matrix $F^{-1}$ 
  \eqn\pent{
  {c_{h_s}(\b_1)  C(\b_1 -h_s b, \b_2,\b_3)\over
  c_{h_t}(\b_2)\, C(\b_2-h_t b, \b_1, \b_3)}
 F_{\b_1 -h_s b, \b_2 -h_t b}= (F^{-1})_{h_t , h_s} \,.
 }
 It is furthermore required that 
 \eqn\penta{
 (F^{-1})_{h_t , h_s}= F_{h_t, h_s}(\b_2,\b_1), 
 }
 a consequence of the pentagon relation for $F$ (or of the normalization relation 
 in \brai{}).  In a shorthand notation 
we shall denote 
$F_{\beta_1-s\o_2 b,
\beta_2-t\o_2 b}(\b_1,\b_2)=F_{s,t}(\b_1,\b_2)= F_{s,t}\,, \ s,t=\pm 1\,, \ 
 F_{\beta_1-\bh b, \beta_2-t\o_2 b}=F_{\bh,t }\,,$ for $\bh:=\o_2-\a_2\,,$ etc.,  
suppressing the dependence on the third argument $\b_3$.

The ratios in \pent\  will  be denoted 
 \eqn\ratoth{
U_{h,h'}(\b_1,\b_2):= {c_{h}(\b_1)  C(\b_1 -h b, \b_2,\b_3)\over
 c_{h'}(\b_2)\, C(\b_2-h'b, \b_1, \b_3)} ={U_{h,+}(\b_1,\b_2)\over  U_{+,+}(\b_1,\b_2) U_{h',+}(\b_2,\b_1)}\,
 }
 and thus one needs to compute all $U_{h,+}$.
We give the explicit expression of the first of these ratios, computed from \formT\
   \eqna\otnc$$\eqalignno{
&U_{+,+}(\b_1, \b_2):={C(\b_1-b\o_2, \b_2, \b_3)\over C(\b_1, \b_2 -b\o_2, \b_3)}=
{
\g(1+b(\b_2-\r Q,\a_2))\g(1+b(\b_2- 2\r Q,\a_2)) \over
 \g(1+b(\b_1-\r Q,\a_2))\g(1+b(\b_1-2 \r Q,\a_2))}
\times \cr
& {\g\(b((\b_{13}^2,\o_1)-{b\over 2})\)\,\g\(b((\b_{13}^2,\o_1)-Q-{b\over 2})\)\over 
\g\(b((\b_{23}^1,\o_1)-{b\over 2})\) \, \g\(b((\b_{23}^1,\o_1)-Q-{b\over 2})\)} \,. &\otnc{}
}$$
By  analogy with 
the Liouville case  this suggests  the following   ansatz for $F$:
  \eqna\fpp $$\eqalignno{
 &  F_{\beta_1-\o_2 b,
\beta_2-\o_2 b}\left[\matrix{\beta_1&\beta_3\cr -\o_2 b&\b_2} \right]=F_{+,+}(\b_1,\b_2)= 
 {\zG(b(Q\r-\zb_2, \a_2))\, \zG(1-b(\r Q-\zb_1, \a_2))\,
\over 
\zG(b((\zb_{31}^2, \o_1)-{b\over 2}))\, \zG(1-b((\zb_{23}^1,\o_1)-{b\over 2})) }\times 
 \cr
 &
{  \zG(b(2\r Q-\zb_2, \a_2))\, \zG(1+b(\zb_1-2\r Q, \a_2))
\over
\zG(b((\zb_{31}^2, \o_1)-{b\over 2}-Q))\, \zG(1-b((\zb_{23}^1,\o_1)-{b\over 2}-Q))\, }\,. &\fpp{}\cr
 }$$
From  \fpp{} one computes  $F_{+,+}(\b_2,\b_1)$ and confirms, using \pent,   that it 
indeed satisfies \penta\
$$
F_{+,+}(\b_2,\b_1) 
=  {C(\b_1
-\o_2 b,\b_2, \b_3)\over C(\b_1,\b_2 -\o_2 b, \b_3)} \, F_{+,+}(\b_1,\b_2) =( F^{-1})_{+,+} \,.
$$
Altogether we have 
 \eqna\fcpp $$\eqalignno{
& U_{+,+}\,F^2_{+,+}= F_{+,+}(\b_2,\b_1) F_{+,+}(\b_1,\b_2) 
=
 {\sin \pi b((\zb_{31}^2, \o_1)-{b\over 2}))\sin \pi b((\zb_{23}^1, \o_1)-{b\over 2}))\over
\sin \pi b(\r Q-\zb_1, \a_2) \, \sin \pi b (Q\r-\zb_2, \a_2)\, 
} \times \cr
&&\fcpp{}\cr
&
{ 
 \sin \pi b((\zb_{31}^2, \o_1)-{b\over 2}-Q))\,  \sin \pi b((\zb_{23}^1,\o_1)-{b\over 2}-Q))\, 
 \over  \sin \pi b(2 \r Q-\zb_1, \a_2)\,  \sin \pi b(2\r Q-\zb_2, \a_2)
 } 
   =: {A B A' B' \over P1[\b_1]P1[\b_2] P2[\b_1] P2[\b_2]}\,.
 }$$
Here $Pk[\b]:=\sin \pi  b ((\b,\a_2)-k Q)  $ 
and  $A,B,A',B'$ denote the four sin's in the numerator correspondingly.

We proceed in this way  to   obtain  $F_{h, +}$ for the other two shifts  of $\b_1\to \b_1-h b$. Then with the help of simple trigonometric relations one  checks and proves 
the first of the diagonal equations in \loceq, for $h_t=h_u= \o_2 $. Similarly
one finds  eight  of the nine $F$ matrix elements checking the related equations. 
The expression for $U_{\bh,\bh}$, however, has  a different structure, 
not suggesting straightforwardly an  
expression for  $F_{\bh,\bh}$

\eqn\tloc{\eqalign{
&U_{\bh,\bh}(\b_1,\b_2)
={\g(3\r Q-b(\b_1,\a_2))\over   \g(1+b(\r Q-\b_1,\a_2))} { \g(1+b(\b_2-3\r Q,\a_2))\over \g(b(\b_2-\r Q,\a_2))}\,.
}}
 On the other hand writing the general expression of an inverse of a $3\times  3$ matrix,
 $ 
  F^{-1}_{ij}
= {\varepsilon_{ikl}\varepsilon_{jmn}\over  2\det F}  F_{mk} F_{nl}
$
with   $i,j,k,l,m,n = +\,,- \,, \bh$ we have, e.g.,  
$$
F_{\bh, +}(\b_2,\b_1)={1\over \det F}(F_{-,+} F_{\bh,-}- F_{-,- }F_{\bh,+})
$$
etc. From this we can determine $\det F$:
\eqn\detf{
\det F(\b_1,\b_2)=
\!=\!-\!\prod_{\a=\a_2,\a_{24}, \a_{14}}{(\zb_1-\r Q, \a)\over (\zb_2-\r Q, \a)}\,.
 }
Then, e.g.,  from
$$
F_{\bh, \bh}(\b_2,\b_1)={1\over \det F} 
(F_{+,+}F_{-,-}-F_{+,-}F_{-,+})
$$
we can determine  $F_{\bh,\bh}$ and check the  remaining identities 
 in \loceq.
\medskip

\noindent
\rb 
$\ $ 
 Summarising we get for the matrix elements of $F$ starting with \fpp{}
  \eqna\fppa $$\eqalignno{
& F_{-,+}(\b_1,\b_2)= F_{+,+}(w_{2132}\bdot \b_1\,, \b_2)\cr
&F_{\bh, +}(\b_1,\b_2) =  F_{\beta_1-\bh b,
\beta_2-\o_2 b}\left[\matrix{\beta_1&\beta_3\cr -\o_2 b&\b_2} \right]={ \G(1-Qb) \over \G(1-2Q b)}\times \cr
&
 {\zG(1+b(\zb_1-3\r Q, \a_2))\,\zG(1-b(\zb_1-\r Q, \a_2))
\,\zG(b((Q\r-\zb_2, \a_2))\, \zG(b((2\r Q-\zb_2, \a_2)) 
\over 
\zG(b((\zb_{13}^2, \o_1)\!-\!{b\over 2}\!-\!Q))\zG(1\!-\!b((\zb_{23}^1,\o_1)\!-\!{b\over 2})) 
\zG(1\!-\!b((\zb_{12}^3, \o_1)\!-\!{\!b\over 2}))\, \zG(1\!-\!b((\zb_{123},\o_1)\!-\!{b\over 2}\!-\!2Q)) }
\cr
& F_{+,-}(\b_1,\b_2)= F_{+,+}(\b_1, w_{2132}\bdot \b_2)
\cr
& F_{-,-}(\b_1,\b_2)= F_{+,+}(w_{2132}\bdot \b_1\,, w_{2132}\bdot \b_2)
&\fppa{}\cr
& F_{\bh, -}(\b_1,\b_2)= F_{\bh, +}(\b_1, w_{2132}\bdot \b_2)\cr
& F_{+,\bh}(\b_1,\b_2) = F_{\beta_1\!-\!\o_2 b ,
\beta_2\!-\!\bh b}\left[\matrix{\beta_1&\beta_3\cr -\o_2 b&\b_2} \right]={ \G(Qb) \over \G(2Q b)} \times  \cr
& {\zG(1-b (Q\r-\b_1, \a_2))\, \zG(1-b((2\r Q-\zb_1, \a_2))\,\zG(b(3\r Q-\b_2, \a_2))\,  \zG(b(\zb_2-\r Q, \a_2)) \over 
\zG(b((\zb_{12}^3, \o_1)\!-\!{b\over 2}))\, \zG(b((\zb_{123},\o_1)\!-\!{\!b\over 2}-2Q)) \zG(b((\zb_{13}^2, \o_1)\!-{b\over 2}))\, \zG(1-b((\zb_{23}^1,\o_1)-{b\over 2}-Q)) }
\cr
& F_{-,\bh}(\b_1,\b_2)= F_{+,\bh}(w_{2132}\bdot \b_1, \b_2)\cr
& F_{\bh,\bh}(\b_1,\b_2)= { \zG((\b_2- \r Q, \a_2)b) \zG(1+(\r Q-\b_1, \a_2)b)\over  \zG(1+(\b_2-3 \r Q, \a_2)b)\zG((3 \r Q-\b_1, \a_2)b)} \times \cr
&\( 1+
{2 \cos \pi Qb\, \sin \pi b ((\b_{12}^3,\o_1) -{b\over 2}-Q)\, \sin \pi b ((\b_{123},\o_1) -{b\over 2}-3Q)\, 
\over \sin \pi b((\b_1,\a_2)-3 Q) \ \sin \pi b((\b_2,\a_2)-3 Q) } \)\,.
}$$
The last  matrix element  can be written in various different ways.

Let us introduce some additional  notation 
 \eqn\notsin{\eqalign{
& D=D[\b_1,\b_2, \b_3] :=\sin \pi b((\b_{123}-2\r Q, \o_1) - {b\over 2})\,, \cr 
&D'
=\sin \pi b((\b_{123}-2\r Q, \o_1) - {b\over 2} + Q)=
D[\b_1,\b_2, w_{13}\bdot\b_3]\,, \cr 
&C 
=\sin \pi b((\b_{12}^3, \o_1) - {b\over 2})=
D'[\b_1,\b_2, w_{2132}\bdot\b_3] \,, \cr 
&C'=
\sin \pi b((\b_{12}^3, \o_1) - {b\over 2}-Q)=
D[\b_1,\b_2, w_{2132}\bdot\b_3] \,, \cr 
}}
and $A, A', B , B'$,  explicitly described above, are also written in terms of Weyl group action 
\eqn\abab{\eqalign{
&\ A=
D'[\b_1,w_{2132}\bdot\b_2, \b_3] \,, A'= D[\b_1,w_{2132}\bdot\b_2, \b_3]\,,  \cr
&B=
D'[w_{2132}\bdot\b_1, \b_2, \b_3] \,,
B'= D[w_{2132}\bdot\b_1, \b_2, \b_3] \,. 
 }}

\noindent
\rb 
$\ $ 
 Denoting $\tF=F(\b_2,\b_1)$,  
 we have  for the products $\tF_{h_s, h_t} F_{h_t,h_s}$ of matrix elements in \loceq\ 
 \eqn\sinid{\eqalign{
 &\tF_{++}F_{+,+}\!=\!{AA'BB'\over P1[\b_1]P2[\b_1]P1[\b_2]P2[\b_2]}\,, \, 
\tF_{+,-}F_{-,+}\!=\!{CC'DD'\over P3[\b_1]P2[\b_1]P1[\b_2]P2[\b_2]}\,,
\cr
 &\tF_{+,\bh}F_{\bh,+}\!=\!{2\cos \pi b^2\,  A'B C D'\over P1[\b_1]P3[\b_1]P1[\b_2]P2[\b_2]}\,,
 \cr
 &\tF_{-,+}F_{+,-}\!=\!{CC'DD'\over P1[\b_1]P2[\b_1]P3[\b_2]P2[\b_2]}\,, \, 
\tF_{-,-}F_{-,-}=  {AA'BB'\over P3[\b_1]P2[\b_1]P3[\b_2]P2[\b_2]}\,,
\cr
 &\tF_{-,\bh}F_{\bh,-}\!=\!{2\cos \pi b^2\,  AB' C' D\over P1[\b_1]P3[\b_1]P3[\b_2]P2[\b_2]}\,,
  \cr
 &\tF_{\bh,+}F_{+,\bh}\!=\!{2\cos \pi b^2\,  AB' C D'\over P1[\b_1]P2[\b_1]P1[\b_2]P3[\b_2]}\,, \, 
\tF_{\bh,-}F_{-,\bh}\!=\!{2\cos \pi b^2\, A'B C' D\over P3[\b_1]P2[\b_1]P1[\b_2]P3[\b_2]}\,, 
\cr
 &\tF_{\bh,\bh}F_{\bh,\bh}=
{ (AA'BB'-CC'DD')^2\over  P2[\b_1]P2[\b_2]
 \prod_{k=1}^3  Pk[\b_1]Pk[\b_2]}\cr 
&= {(\cos \pi b^2 \,  \cos \pi  b((\b_3,\a_2) -2 b) -\cos \pi b( (\b_1,\a_2) -2 b)  \cos \pi b((\b_2,\a_2) -2 b) )^2
 \over  P1[\b_1]P1[\b_2] P3[\b_1]P3[\b_2]}\,.\cr
  }}
 Compare  with the  Liouville case where  $
  (\b, \a) =2\b^L \,, (\b, \o) =\b^L$ and 
 $$
  F^L_{s, t}= F^L_{\beta_1-s  \o b, \beta_2-t \o b}\left[\matrix{\beta_1&\beta_3\cr
-\o b&\beta_2} \right]\,, \, \  s,t=\pm 1 
 $$
satisfying\foot{
Analogous  relations hold in the $sl(2)$  WZW  case. The braiding matrices    differ by $\beta$ - independent phases (Q is replaced by $b$) and by normalisation so that effectively  $\det F=1=\det B$, i.e., $F\in SL(2)$.  }  
 
\eqn\liof{\eqalign{
&  F^L_{+,+}\tF^L_{+,+} = F^L_{+,+}{1\over \det F}
 F^L_{-,-}=\tF^L_{-,-}F^L_{-,-} ={ A B\over P1[\b_1] P1[\b_2]}\,,
\cr
&  F^L_{+,-}\tF^L_{-,+} = - F^L_{+,-}{1\over \det F} F^L_{-,+}= \tF^L_{+,-}F^L_{-,+} ={ C D \over P1[\b_1] P1[\b_2]}\,.
 }}

One can analogously compute  the fusing matrix elements corresponding to the $sl(4)$ fundamental  weights $\o_1=\o_3^*$ using the 3-point constant computed by Fateev and Litvinov 
\FLc\ in which two of the  weights are  arbitrary and the third is proportional to one of these fundamental weights; this will be a special case  of the $4 \times 4 $ F matrix.   Partial data  on the braiding matrices in that  case is also provided (though in a different gauge) by the Boltzmann weights defining 
 integrable $A_3^{(1)}$ lattice models \JMO\ taking a proper limit of the spectral parameter.

\bigskip

\noindent
\rb 
$\ $ Finally  we   check a  braiding relation
 relevant for the 4-point chiral blocks  under consideration, 
namely (choosing  the sign  $\epsilon=1$ in \brai{})
\eqn\eqms{
\Omega_1 \Omega_2\Omega_3:= (e_1^2) (e_2 e_1^2 e_2^{-1}) (e_3 e_2 e_1^2 e_2^{-1}e_3^{-1})= e^{-4\pi i \triangle(f)}\,.
}
 In the limit $b\to 0$  the r.h.s of \eqms\ becomes an identity  for any of the fundamental weights $f=-\o_i b$. In our case $\tri(f)=\tri(-\o_2 b) =-{5\over 2} b^2-2$.

The meaning of the l.h.s  of \eqms\ is  a composition  of monodromies  around the  three vertex coordinates. 
On the sphere with 4 (ordered) points $e_1$ and $e_3$ are represented by diagonal braiding matrices;
$e_2$ in $\Omega_2$  is represented  by \brai{}, while in 
$\Omega_3$ it is represented by the same braiding matrix with $\b_3$  and $\b_1$ exchanged.
Using the defining relations of the braiding group
\eqn\braidrel{
e_i e_{i+1} e_i= e_{i+1} e_i e_{i+1}\,,\  e_ ie_j=e_j e_i \ {\rm for}\ j\ne i\pm 1
}
\eqms\ is reduced to the first of the two additional relations on the generators $\{e_i\}$
which characterise the modular group on the sphere with 4 holes \MS
\eqn\ms{e_1 e_2 e_3^2 e_2 e_1= e^{-4\pi i \triangle(f)}\,, \ (e_1e_2e_3)^4=
e^{-2\pi i (\triangle(f)+ \sum_{a=1}^3 \tri(\b_a)) }\,.
}
 Take  the trace of \eqms
\eqn\eqmstr{
{\rm Tr}(\Omega_1 \Omega_2)=e^{-4\pi i \triangle(-\o_2 b)}{\rm Tr}( \Omega_3^{-1}) \,.
}
The eigenvalues of the monodromy $\Omega$ are computed from the difference of Toda dimensions
\eqn\eigvtoda{
p(\beta, h)= \triangle(\b - hb)-\triangle(\b)-\triangle(-b\o_2)= 2bQ+ b (\b-\r Q, h )
}
where in general $h\in \Gamma_{\o_2}$ is a weight in the 6-dim weight diagram $\G_{\o_2}$ of the fundamental representation $\o_2=(0,1,0)$,  i.e., $h=\pm \o_2, \pm (\o_2-\o_1-\o_3), \pm (\o_1-\o_3)$. Thus  in general the trace of the diagonal monodromy 
\eqn\montoda{
{\rm Tr}(\Omega)={\rm Tr}(e_1^{2})= \sum_{h\in \Gamma_{\o_2}}  e^{2\pi i p(\b, h)}=
e^{i\pi 4bQ} \chi_{\o_2}(2\pi i  b (\b-\r Q))
}
is proportional to  the character $\chi_{\o_2}(\mu)$  of the fundamental representation $\o_2=(0,1,0)$ evaluated at the "angle"
$\mu=2\pi   b (\b-\r Q)$. (For $f =\o_i$ in the formula corresponding to \eigvtoda\  $h\in \Gamma_{\o_i}$ and 
the constant term in the r.h.s.  is given by $( \o_i, \r) b Q$.) Denote by $q(\b)$ the normalized diagonal matrix 
$$q_{h_s,h_t}(\b)=\delta_{h_s,h_t}\, e^{2\pi i b(\b\!-\!\r Q, h_s)}=e^{-4\pi i Q } \Omega\,.
$$
 In terms of
 $F$ and its inverse $\tF$ the relation \eqmstr\  reads (collecting the three overall  terms $e^{4 i \pi b^2}  $ in the r.h.s.)
 \eqn\janid{
\sum_{h_s,h_t} q_{h_s}(\b_1) F_{h_s, h_t}\tF_{h_t,h_s}q_{h_t}(\b_2)=e^{-4 i \pi (3 b^2 +\triangle(-\o_2 b))}
{\rm Tr}\,  q^{-1}(\b_3)\,.
 }
In our case  of "scalar" (in 4d sense)  weights $\b_a$ the relation involves a  $3\times 3$ submatrix  and accordingly 
$q(\b)$ reduces to  the diagonal  submatrix with matrix elements $e^{2\pi i(\b-\r Q, h_s)}\,, h_s=\pm \o_2\,, \o_2-\a_2$. Thus the sums in the l.h.s. of \janid\  run over these three weights, while  the r.h.s.  reduces to 
\eqn\redjan{
{\rm r.h.s.}'\,  =e^{-2 i\pi b^2} (2 \cos 2\pi  b(\b-2\r Q, \a_2) +e^{2\pi i b^2})\,.
}
Each of the products in \sinid\  which appear in the  l.h.s. of \janid\  in this case  is a second order polynomial in  $(2 \cos \pi  b(\b-2\r Q, \a_2) $, as is the  expression  in \redjan, so the reduced relation is checked  
  order by order.  

In 
\janid\  $F$  enters only  through  the products  $F_{h_s, h_t} \tF_{h_t,h_s}$, hence  
it  is a restriction on these products. 
In principle the identity \eqms\ 
with diagonal braiding determined from
\eigvtoda\  may 
admit more general solutions  for the individual $F$ matrix elements 
 than the present Toda  CFT solution
\fpp{}, \fppa{}. 
Indeed the $sl(2)$  analog of  the identity \eqms\ with trivial r.h.s.  has been 
exploited in the recent papers \JW, \KK\  on AdS$_3 \times S^3$ sigma model 3-point correlators in the semi-classical  strong 't Hooft coupling $\l$ limit  with large quantum numbers. Identifying 
$b^2=1/\sqrt{\lambda}$ this corresponds to the  semiclassical limit $b\to0$ with three heavy charges  $\beta_a/b= \eta_a/b^2 $, $\eta_a $ - finite. In the sigma model case 
 the  eigenvalues of the monodromy matrix  $e^{ 2\pi i (\eta(x),h)}$ depend on the spectral parameter $x$ and the solution for the individual $F=F(x)$ matrix elements depends  nontrivially on the specific spectral curve. On the other hand   the expression for the products 
$F_{h_s, h_t} \tF_{h_t,h_s}$
 as functions of $\eta(x)$ 
coincides with  those in  the WZW model, or,  up to normalization, with  those of Virasoro theory
(cf.  \liof\ and footnote 4). One may expect that the Toda theory data  and their WZW extensions for the fundamental representations $f=-\o_i b$ can similarly be used as a starting point, although in this case the equation  \eqmstr\ is less restrictive by itself, compared with the sl(2) case where it  uniquely determines the  fusing matrix products.\foot{ 
 The  implicit assumption in \JW,\KK\   concerning  the identity  \eqms\  along with the hexagon identities 
 \braidrel\ is equivalent to the statement that the 
  2d four-point  solutions (normalized by the 3-point function) of the  auxiliary linear   system of equations 
 transform  linearly with respect to the $F(x)$ matrix, so that   the Wronskians of two related in this way solutions 
 are expressed by $F(x)$
matrix elements.}

\newsec{ The 3-point functions in the compact  ("matter") region and  BPS - like relation}

By analogy with the Liouville gravity described by two dual Virasoro CFT with
$c>25$ (Liouville) and $c<1$ ("matter") we shall extend here the results of section 2 
to  another region of central charge of the $W_4$ CFT, parametrised by the same real 
 parameter $b$ as \tod, 
\eqn\mat{
c_{m}= 3 (1-20 e_0^2)  
<3 \,,  \ e_0={1\over b}-b \,.
}
The sum of central charges \tod\ and \mat\ is compensated by the contribution of the ghosts
(pairs $(b_k,c_k)$ of dimensions  $ (k, 1-k)\,, k=2,3,4$), i.e.,   $c+c_m+c_{\rm ghost}=0$.
 
The   conformal dimension of vertex operator $V_e^{(\rm m)}$ is given by
 \eqn\dimmat{
\tri_m(e)= {1\over 2}(e, e-2\r e_0)\,, 
 }
 invariant under the action of the Weyl group
 \eqn\weylmat{
 \hat{w}(e):= \rho e_0 +w(e-\rho e_0)=b( w\cdot {e\over b} -{1\over b^2} w\cdot 0)\,
 }
(i.e.,  the  horizontal projection of  the shifted action of the affine Weyl  group elements 
 $t_{-w.0}w $ on $(e/b +k\o_0)$,  times  $b$, where  $k+4= 1/b^2$). 
 
 The minimal $W_4$ theory  in the region \mat\ for rational $b^2$ has been discussed in 
 \FLuk. Here the real parameter $b$ is generic  and we shall consider
vertex operators $V_e^{(\rm m)}$  with symmetric charges $e= (r \o_2 +s(\o_1+\o_3)) b=e^*$. Such  $W_4$ representations 
are degenerate for  nonnegative integers $r, s$. 
Once again we consider   a 3-point function of vertex operators two of which have highest weights of type   $e_a=r_a \o_2 b\,, a=1,2$, 
and one - a  general symmetric weight  $e_3=e_3^*$. 
The Coulomb gas computation is  performed as before, with interaction term defined by vertex operators $V_{-\a_i b}^{(\rm m)}$,  or,   one can directly continue the  Toda Coulomb gas OPE constants (being given by finite  products of ratios of  $\gamma $ functions) to  
  $b^2\to  -b^2\,, Qb\to e_0b \,,  \b b \to e b$. 
  This OPE constant   can be expressed directly
 in terms of  $\zu$-functions with the result
  \eqna\opem $$\eqalignno{
& C_m(e_1, e_2, e_3)= R_m(e_3)\,  C_m(e_1, e_2, 2\r e_0-e_3)\cr
&= {\(b^{2Qb}\l_m\)^{ {(e_{123}-2\r e_0,\r)\over b}}} 
{\prod_{\a=\a_1,\a_{13}}\zu((e_3-\r e_0,\a)+b)\over \prod_{\a=\a_1,\a_{14}}\zu((e_3-\r e_0,\a)+e_0 +b)}\times \cr
 &\prod_{a=1,2} {\zu((e_{123}-2e_a,\o_2-\o_1)+b) \over   \zu((e_a,\a_2)\!+\! b) }{ \zu((e_{12}^3,\o_1)\!+\!b) \zu((e_{123}\!,\o_1)\!-2e_0+\!b)   \over  \zu(b)  \zu((e_3,\a_2)\!+\! b) }\times &\opem{}\cr
 &
 \prod_{a=1,2} {  \zu((e_{123} -2e_a,\o_1)-e_0+b) \over  \zu((e_a,\a_2)\!-\!e_0\!+b)}\,  { \zu((e_{12}^3,\o_2\!-\!\o_1)\!-\!e_0\!+\!b) \zu((e_{123},\o_2-\o_1)\!-\!3e_0\!+\!b)  \over  \zu(b)  \zu((e_3,\a_{24}) \!-\!e_0+b)}
 }$$
 where $\l_m= \pi \mu_m \gamma(-b^2)$ with $\mu_m$ - the analog of the cosmological constant, multiplying the interaction term in the action. The reflection amplitude corresponding to the longest Weyl group element 
 $w_{121321} $ is the analytic continuation of  \reflampl\  (written first as a finite ratio of $\g$-functions and then rewritten in terms of $\zu$-functions)
 \eqn\weilmat{
 R_m(e_3)= \(b^{2Qb}\l_m\)^{ {(2 e_{3}-2\r e_0,\r)\over b}} \prod_{\a>0}
 {\zu((e_3-\r e_0,\a)+b)\over \zu((e_3-\r e_0,\a)+e_0 +b)}\,.
 }
Analogously to \weylrep\  the eight  three charge  factors  in   \opem{} can be written as points on an orbit with respect to  the shifted Weyl action \weylmat. 
The $F$- matrix elements  are obtained by the same  analytic continuation
of the Toda ones in \fpp{},   e.g., 
  \eqna\fppmat $$\eqalignno{
 &  F^m_{e_1+\o_2 b,
e_2+\o_2 b}\left[\matrix{e_1&e_3\cr \o_2 b&e_2} \right]=
 {\zG(b(e_0\r-e_2, \a_2))\, \zG(1-b(\r e_0- e_1, \a_2))\,
\over 
\zG(b((e_{31}^2, \o_1)+{b\over 2}))\, \zG(1-b((e_{23}^1,\o_1)+{b\over 2})) }\times 
 \cr
 &
{  \zG(b(2\r e_0- e_2, \a_2))\, \zG(1+b(e_1-2\r e_0, \a_2))
\over
\zG(b((e_{31}^2, \o_1)+{b\over 2}-e_0))\, \zG(1-b((e_{23}^1,\o_1)+{b\over 2}-e_0))\, } &\fppmat{}\cr
 }$$
etc..

\medskip

\noindent 
\rb
$\ $ The $W_4$ CFT is described alternatively as the (principal)  quantum 
DS  reduction
of a  $\hat{sl}(4)$  WZW model (or its dual). 
With the parametrisation in \tod\ and \mat\ in the noncompact and compact WZW analogs  
 the corresponding Sugawara dimensions are given by

 \eqn\sugaw{ 
 \tri^{\rm Su}(\b)= {1\over 2}(\b, 2\r b -\b)\,, \ \  
\tri^{\rm Su}_m(e)= {1\over 2}(e, e+2\r  b)\,, 
 }
 invariant (along with the higher Casimir eigenvalues) under the standard shifted action of the Weyl group
 on the  $sl(4)$ weights $-\b/b$ and $e/b$. The dimensions of the vertex operators in the WZW theory and their reduced Toda counterparts   are related as\foot{On the level of 2-and 3-point functions the reduction amounts (up to a constant)  to a $"x\to z"$ limit of the $sl(4)$  isospin variables, see \FGPP\ and references therein. In particular,    for the vertex highest weights of type $ (\l,\a_1)=0= (\l,\a_3)\,, $ to which we shall restrict in what follows, they are   described by a 4d vector $x_{\mu}$  and  the  $"x\to z"$  limit  reads $x_{ij}^2  \to   
 |z_{ij}|^4$. E.g.,  applied to the  
 WZW 2-point functions 
 $ G^{(m)}_{e}(x_{12}; z_{12}, \bar{z}_{12})=  (x_{12}^2)^{(e,\r)/2b} \, |z_{12}|^{-2\tri_m^{\rm Su}(e)}$
 and 
  $ G_{\b}(y_{12}; z_{12}, \bar{z}_{12}))=  (y_{12}^2)^{-(\b,\r)/2b} \, |z_{12}|^{-2\tri^{\rm Su}(\b)}$
this  reproduces,  in agreement with (4.8),   the corresponding   2-point functions of the $W_4$ fields  up to  constants.}
 \eqn\red{
 \tri(\b)= \tri^{\rm Su}(\b)+ {1\over  b }(\b, {\r })\,, \ \ \tri_m(e)= \tri_m^{\rm Su}(e)- {1\over  b}(e, {\r })\,.
}
For any pair of weights $\b$ and $e$ related by an element $w$ of the Weyl group one has a  BPS-like relation
\eqn\bps{\eqalign{
&\b = -b\, w\cdot {e\over b} = b\r - w (e +b \r) \cr
&\Rightarrow \ \ \tri^{\rm Su}(\b)+
  \tri^{\rm Su}_m(e)=0\,.
}}
In particular there is only one  nontrivial element of Weyl group, $w_{2132}$,   s.t. its shifted action 
 preserves the 
$sl(4)$ representations of type   $ (\l,\a_1)=0=(\l,\a_3)$,  namely, 
$w_{2132}\cdot \l=-\l-4\o_2$,  so that the first line of \bps\ reads   
\eqn\exbps{
\b=-b  w_{2132}\cdot {e\over b}=e + 4\o_2 b\, \ \Rightarrow   (\b, \a_2)/b=(e,\a_2)/b+4\,.
}
While the sum of Sugawara dimensions vanishes according to \bps, for the sum of the related   by  \red\ $W_4$ dimensions one has $\tri(e +4b\o_2)\!+\!\tri_m(e)\!=\!8$.

Recall that in  the $sl(2)$  case the relation in the first line of \bps\  and its dual
 yield for the Virasoro dimension  $\tri(\epsilon e +\a b^{\epsilon})\!+\!\tri_m(e)\!=\!1\,, \epsilon=\pm 1$. Accordingly the products $c \bar{c}\,  V_\b V_e^{({\rm m})}$ (where $c, \bar{c}$ are the chiral components of the  ghost   of dimension $-1$),  or  $\int d^2x V_\b V_e^{({\rm m})}$,   describe BRST invariant operators  - the tachyons of the Liouville gravity.  
They  have trivial,  up to leg factors,  3-point function \AlZ,\KPa.  Apparently unlike the Virasoro case  one cannot realise $W_4$ analogs 
of such operators
through  products of vertex operators from the two regions of the theory.

Nevertheless in  view of the
relation between  the $ \hat{sl}(4)$ WZW and the $W_4$ conformal theories
we may expect that 
the 3-point constants in the two $W_4$ regions are closely  related. Indeed, take 
all $e_a=(0,r_a,0) b$ and impose  \exbps,  
i.e., $ C(\b_1,\b_2,\b_3)=C(e_1+4\o_2 b,e_2+4\o_2 b,e_3+4\o_2 b)$. One then has for the product of the two related constants
\eqna\mattod $$\eqalignno{
& C_m(e_1, e_2, e_3) \,  C(\b_1,\b_2,\b_3) =\prod_{a=1}^3 \phi((\b_a,\a_2))\, A(\b_1,\b_2,\b_3)\, 
\bar{C}_m(e_1, e_2, e_3) \,  \bar{C}(\b_1,\b_2,\b_3)\, \cr
&=\l_T^{(2\r Q- \b_{123}, \rho)\over b}\l_m^{(e_{123}-2\r e_0, \rho)\over b}
 \prod_{a=1}^3 {(b^2)^{3} \prod_{\a=\a_2,\a_{24},\a_{14} } \g((\b_a-\r b, \a)b)\ \over  
\g(( \r Q -\b_a, \a_{24})b) }\, A(\b_1,\b_2,\b_3)\,  
\,, \cr
&{\rm where}\ &\mattod{}\cr
&\  A(\b_1,\b_2,\b_3)=  
  \((1-b^4)^2 
  ((\b_{123}-2\r Q, \o_1)b)^2 \prod_{a=1}^3   ((\b_{123} - 2\b_a, \o_1) - Q)b)^2 \)^{-1}\,.\cr
}$$
\medskip
\noindent
 The $\g$-factors in the second line of \mattod{} (analogs of the leg factors in Liouville  gravity)  can always be removed  by proper field normalisation. The intermediate notation $\bar{C}$ and $\bar{C}_m$
 in the r.h.s. of the first equality refers to the constants obtained from the corresponding $C$ in \formT{} and $C_m$ in \opem{}  by replacing $Q\to b$ and $e_o \to -b$, respectively, in the $\zu$-functions. This is  achieved by the use of one of the functional relations \fnct\ and  produces  finite products of $\gamma$-functions for each of the two constants,  
  that  are furthermore compensated in the product $CC_m$ up to the factor $A(\b_1,\b_2,\b_3)$ 
in \mattod{}  and $\prod_a \phi((\b_a,\a_2))$, the explicit expression of which we skip.  As clear from \mattod{}   the  product $\bar{C}\bar{C}_m$ itself  is trivial up to field renormalisation: 
 the modified denominator  from  
  the third (fourth) line in \formT{}  cancels the modified numerator  from  the fourth (third) line in \opem{} respectively.

One may expect that  the two 
   constants  $\bar{C}_m(e_1,e_2,e_3)$ and $\bar{C}(\b_1,\b_2,\b_3)$ 
will describe 
the corresponding 3-point constants 
of the compact and noncompact  WZW model.
This conjecture remains to be checked. In any case  the triviality 
of the product $\bar{C}\, \bar{C}_m$  whenever  
the  relation  \exbps\  is imposed is a property 
    expected 
  for the correlators of BRST invariant states in the non-critical string model described by  a   $G/G $ topological CFT, see, e.g., \ASY.

In  the  semi-classical limit $b\to 0$ with  "light"  charges, i.e., $(\b_a, \a_2)/b=\sigma_a$
 are assumed  finite,  
 the factor  in \mattod{} which depends nontrivially on the three charges  goes to a numerical constant, $A(\s_1 b,\s_2 b,\s_3 b)\to 1/9$. In other  words in this limit the cancellation expected for the WZW counterparts of the  $W_4$ constants holds true for the Toda  constants themselves.

We conclude with a remark about  the  "light-charge"   limit of each of the constants   $\bar{C}(\b_1,\b_2,\b_3)$  and $\bar{C}_m(e_1,e_2,e_3)$ 
computed using  
 the asymptotics of $\zu(x)$
$$\lim_{b\to 0}\zu( b)/\zu(\s b)=\Gamma(\sigma)b^{\sigma-1}\,.$$ 
As explained above   in these constants  compared to the initial Toda ones one replaces $Q\to b$ and $e_0\to -b$. All weights are taken to be proportional to the second fundamental weight $\o_2$, $\b_a=\sigma_a \o_2 b\,, e_a=r_a\o_2 b$. We have in the limit $b\to 0$ with finite $\sigma_a\,, r_a$
\eqna\cllm$$\eqalignno{
\bar{ C}(\b_1, \b_2, \b_3 )    \sim &\G({\s_{123}\over 2}- 2) 
\prod_{a}{\G({\s_{123} \over 2} - \s_a) \over \G(\s_a)   }\,\times   &\cllm{} \cr
& \G({\s_{123}\over 2}- 3) 
\prod_{a}{\G({\s_{123} \over 2} - \s_a-1) \over \G(\s_a-1)   }
}$$
and 
\eqna\cllmmat$$\eqalignno{
\bar{ C}_m(e_1, e_2, e_3 )  \sim  & {1\over \G({r_{123}\over 2} +3) }
\prod_{a}{\G(r_a+1)  \over \G({r_{123} \over 2} - r_a+1)  }\,\times &\cllmmat{} \cr
& {1\over \G({r_{123}\over 2} +4) }
\prod_{a}{\G(r_a+2)  \over \G({r_{123} \over 2} - r_a+2)  }\,.
}$$

One  recognizes in the 
$\Gamma$-function ratios  of 
the first lines in \cllm{}  and \cllmmat{} 
precisely the 
expressions of  the 3-points constants of scalar 4d fields 
computed by 
 integrating the bulk-bounday  kernels (classical vertex operators) over  the cosets $AdS_5$ and $S^5$, respectively
 \FMMR, \LMRS. In this comparison   we identify the charges 
$(\b_a,\a_2)/b $  - with the 4d scalar field conformal dimensions $\tri_a$ 
and  the weights $e_a/b$ (taking  nonzero integer values) with  the 4d 
isospins given by   the SU(4) representation $(0,J_a,0$).\foot{ These are the  $AdS_5$ and $ S^5$ free field ingredients of the 3-point function of "chiral primary operators"  with $\tri_a=J_a$ 
 at strong coupling $\l$  \LMRS.   
The full correlator involves 
an additional factor coming from  the  coupling constant of the supergravity cubic interaction term,  which compensates   the product of the expressions in the r.h.s. of  the  first lines of  \cllm{}  and \cllmmat{} taken with $\s_a=\tri_a\,, r_a=J_a$
(formula (3.40) of \LMRS).
} 
The 
condition \exbps\ for which  the product of  \cllm{} and \cllmmat{} trivialises  implies with 
such identification $J_a=\tri_a-4$.

On the other hand  we can identify $(\b_a,\a_2)/b =(\b_a,\o_2)/b$  with $\tri_a+4$ instead. Then 
neither of the two  factors 
 in $\bar{C}$  \cllm{}  reproduces the $AdS_5$ result, but the 
trivialisation 
of the full  $\bar{C}\, \bar{C}_m$ (and, in this limit, of  the Toda constants product  $C C_m$ itself) due to \exbps\ holds true for 
$J_a=\tri_a$, which is the actual 4d supersymmetric BPS condition for the given class of representations; the second line in \bps\ is equivalent to the vanishing of the second Casimir of  the superconformal  algebra $sl(2,2|4)$.\foot{Different identifications for the three weights are also possible (reminiscent of the mixed correlators discussed in \AF).} Note that Toda light charge classical correlators can be 
  computed alternatively by integrals of the exponential fields over the  "bulk" $SL(4)$ group, as 
  shown on  examples 
   in \FLc, generalising the 
  computation \ZZ\  in the Liouville case.

\newsec{Concluding remarks.}

We have constructed   3-point functions in the $W_4$ Toda theory and have  used them to derive 
 novel  data on a  fundamental braiding/fusing  matrix extending   the rank $1$ results.  The solution described by a $3\times 3$ matrix applies to a particular class of representations arising from  partially degenerate Verma modules   with highest  weights proportional to the sl(4) fundamental weight $\o_2$. 
The  examples of OPE structure constants 
 computed here and in \FLc\  are  still quite simple
and need to be extended to positive integer "4d spin"  components  $l_a^{(i)}=-(\b_a, \a_i)/b \,, i=1,3$. 
  For that  purpose the  AGT-W approach \MP\   might be more constructive. 
On the other hand one can try to exploit  the pentagon equation for the $6 \times 6$ $F$   matrix
as a recursive relation given the initial data computed here and in \FLc.

We have analysed a higher rank   analog of the  braiding relation which played a basic role in the construction of the semi-classical  limit of  a class of  3-point functions  on  
$AdS_3\times S^3$   \JW,\KK\  and have identified it with a standard identity in the modular group on the plane with four holes. 
The explicit data  for  the solutions of the braiding identity  provided by   Toda CFT, 
in particular their "heavy charge" limit,  may thus find application to 
 the quasiclassics of  conformal  
sigma models described by compact and noncompact  forms of $SL(4,\CC)$, generalising the $SL(2,\CC)$ results. Here again for a realistic application one needs first to extend the result beyond the particular class of representations.

More precisely,  for this application one needs 
 the extension of the Toda  modular data to that of its  
WZW model counterpart; we hope to return to this problem. 
The computation of the corresponding $\hat{sl}(4)$ WZW 3-point functions  is 
  important also 
 in view of the possible application  to the   $G/G$ models. As  we have observed,
  the affine sl(4)  WZW  theories  can  alternatively  describe the simplest 
 BPS states in the    "light charge" classical limit   by a  different mechanism than  the  one  provided by  the 
 supergravity  approximation.
 The  2d CFT expected to describe the  worldsheet realisation  of the $\CN=4$ YM theory   lacks the affine symmetry of the  (super)conformal  WZW models. 
Nevertheless  further development of the latter may provide some  inside on the structure of the former. 
 \bigskip  

\no
{\bf Acknowledgements}
\medskip
\noindent
We  thank 
 Ivan Todorov for  a  useful  discussion concerning \ms.
VBP acknowledges the    hospitality of  the Istituto Nazionale di Fisica Nucleare (INFN),
Sezione di Trieste, Italy. This work  is partially supported by the Bulgarian NSF Grant
DFNI T02/6 and by the COST action MP-1405 QSPACE.

\vfil\eject
\appendix{A}{Details on the calculation  of the Coulomb integrals
}

\noindent
$\bullet \ $  We  start with briefly recalling   the technique \FLc\  of computation of some multiple integrals generalising Selberg integrals.
The Toda 3-point Coulomb integral  (with one type of screening charges) 
is given by 
\eqna\todainteg$$\eqalignno{
&I_{s_1,s_2,s_3}(\b_1,\b_2)
 =
\int  \prod_{k=1}^3 d\mu_{s_k} (t^{(k)})\, D_{s_k}^{-2b^2}(t^{(k)}) \times \cr
 && \todainteg {} \cr
& \prod_{i, j}^{s_1,s_2}\Big(|t_i^{(1)}-t_j^{(2)}|^{2b^2}\Big)\prod_{i, j}^{s_3,s_2}
\Big(|t_i^{(3)}-t_j^{(2)}|^{2b^2} \Big)
\prod_{k=1}^3 \prod_{i=1}^{s_k}  |t_i^{(k)}|^{-2 (\b_1, \a_k)b}|t_i^{(k)}-1|^{ -2  (\b_2, \a_k)b}
 }$$
 where 
 $$D_{s}(t)=\prod_{i<j}^s |t_i-t_j|^{2}\,, \ \
  d\mu_s(t)= {1\over \pi^s s!}\prod_{j=1}^s d^2 t_j\,.
 $$
 The integral can be computed recursively for particular  sets of weights $\b_1, \b_2$,   exploiting 
 a  duality formula  \FBas\  originating from  the Virasoro theory of central charge $c=-2$  
   \eqna\flid$$\eqalignno{
\int d\mu_n(y) D_n(y) \prod_{i=1}^n\prod_{j=1}^{n+m+1}|y_i- t_j|^{2p_j}&=
{\prod_{j=1}^{n+m+1}\g(1+p_j)\over \g(1+n+\sum_j p_j)}
\prod_{i<j}^{n+m+1}|t_i-t_j|^{2+2p_i+2p_j}\times \cr
&& \flid{} \cr
 &\int d\mu_m(u) D_m(u) \prod_{i=1}^m\prod_{j=1}^{n+m+1}|u_i- t_j|^{-2-2p_j}\,.
}$$
This formula  results from   two alternative 
 Coulomb gas representations 
 of the $n+m+2$- point function, obtained 
 by replacing each vertex with its dual of the same conformal dimension; 
 the compatibilty  of the two charge conservation conditions, involving different numbers of screening charges, fixes 
  the parameter $b$ parametrizing the central charge.
The  
 two  integral representations  coincide up to a constant $C_n(\{p_j\})=C^{-1}_m(\{-1-p_j\})$, indicated in the r.h.s. of \flid{}, which is 
 given 
by a product 
of reflection amplitudes. 
\medskip 
\noindent 
$\bullet \ $ For the  particular  integral discussed in section 2  the dependence on the two charges in
\todainteg{}   simplifies since $(\b_1, \a_i)=0=(\b_2, \a_i)$ for $i=1,3$. The  calculation of the  integral  starts applying  \flid{}  for $n=s_1-1\,, m=0\,$ and $p_j=-Qb\,, j=1,\dots , s_1\,, $ identifying the power of coordinate differences  in the r.h.s. with  the  factor  
$D_{s_1}^{1+2p_j}(t^{(1)})=D_{s_1}^{-2b^2}(t^{(1)})/ D_{s_1} (t^{(1)}) $ in  \todainteg{}.    The formula \flid{}  is then applied  to the integrals over $\{t^{(k)}\}, $ sequentially for  $ k=1,2,3$ with $m=s_2-2\,, m=s_3-1\,, m=0\,,$ respectively. The result is an  integral of the same type as 
 \todainteg{},   with shifted indices and arguments described   in section 2.  After $s_1$ steps
 one obtains
 \eqn\recr{\eqalign{
 I_{s_1,s_2,s_3}(\b_1,\b_2)=K I_{0,s_2-2s_1,0}(\b_1+s_1 b \o_2,\b_2+s_1 b \o_2)
 }}
where the integral in the r.h.s is  a Coulomb Liouville  integral  with $s_2-2s_1$ screening charges
\eqn\lioucc{\eqalign{
\sum_{a=1}^3 \b_a^L+(s_2-2s_1) b=Q\,, \  2\b^L_a:= 2(\b_a,\o_1) +s_1 b\,, a=1,2
\,.
}}
It is  a residue of the DOZZ formula at the values corresponding to the charge conservation condition  in \lioucc, or,  in Toda variables - at $(\b_{12}^3, \o_2-2\o_1)=(\b_3, \a_1)
=-l b$ with nonnegative integer $l=s_2-2s_1$.

The  constant $K$ in \recr\  is given by  
  \eqn\formTnew{\eqalign{
&K= 
 \prod_{s=0}^{s_1-1} \(b^{ ((s_1 +1) b^2 +1 )}\, \g((s -s_1) b^2)\)
\lim_{\varepsilon\to 0} { \zu(2Q+(s_2-2s_1)b+\varepsilon) \over  
 \zu(2Q+(s_2-s_1)b+\varepsilon)}
\cr
& ({b^{-2 Q b}\over \g(-b^2)})^{-4(\b_{12}^3, \o_3)\over b}  \, \prod_{\a=\a_2,\a_{24}} \prod_{a=1,2}
 {\zu((Q\r-\b_a,\a)) \over \zu((Q\r-\b_a,\a)-s_1 b)} \cr
&\cr
&
{  \zu(3Q-s_2 b -(\b_1+\b_2, \a_2))   \zu(4Q-2s_1 b -(\b_1+\b_2, \a_2))
\over  
 \zu(3Q+(s_1-s_2)b -(\b_1+\b_2, \a_2))  \zu(4Q-s_1b -(\b_1+\b_2, \a_2))
} \,.
}} 
In writing \formTnew \ we have used the functional relation \fnct\  
 to replace products of $\gamma$ - functions with ratio of $\zu-$ functions. 
 The ratio of (regularized) $\zu$-functions in the r.h.s of 
 the first line is a 
 finite product of $\gamma$'s,  written in a compact form. This factor
   can be rewritten getting  rid of the nonnegative  integers  $s_1,s_2$ using \cconsc\  and then   
can be continued  for arbitrary $\b_a$ of the type in \clasreps\ without the restrictions implied by   \cconsc, thus giving ${\zu((\b_3, \a_1)-Q)\over \zu((\b_{12}^3, \o_2-\o_1)-Q)}$ .  Analogously are rewritten and continued 
the  factors  in the second and the third line. 
On the other hand  the product in the first line can be written as a residue of an  analytically continued expression  
\eqn\resi{{\rm  res}_{(\b_{12}^3, \o_1) = -s_1 b} \,{\Upsilon(b) \over \Upsilon((\b_{12}^3, \o_1) )} 
=b^{s_1[(s_1+1)b^2+1]} \prod_{s=0}^{s_1-1} \gamma ((s-s_1)b^2) \,.
}
Altogether, combining with the Liouville constant discussed above, one obtains the  expression for the OPE constant $C(\b_1,\b_2,2\r Q-\b_3)$ in  \formT{}. It is   valid 
for weights of the type in \clasreps{},  while the Coulomb OPE is recovered as in \resdoubl.

 \bigskip 

\noindent
$\bullet\ $ 
A slightly more general case,  in which the  OPE constant 
$c(\b_1,\b_2,2\r Q-\b_3)$ can be computed along the same path,  is provided by charges $\b_1,\b_2$ s.t., say,   $\b_2$ is of the same kind as before, $(\b_2,\a_i)=0, i=1,3\,,$ while $\b_1$ has two nonvanishing components,  e.g., $  (\b_1,\a_1)=0\,$. 
The integral 
 is computed under the condition $s_3\ge s_1$. After $s_1$ steps the $sl(4)$ type integral reduces to a $ sl(3) $ type $I_{0,s_2-2s_1, s_3-s_1}$  which furthermore is reduced to  Liouville type $I_{0,s_2-s_1-s_3, 0}$. In particular the  resulting expression for the example $\b_2=-\o_2 b\,, 
 s_1=0\,, \, s_3=1=s_2$ reproduces the 
 OPE formula (1.51) in  \FLc\  for the shift
$\b_3=\b_1-(\o_2  -  \a_{24}) b=\b_1-(\o_1-\o_3) b$:
$$
c(\b_1,-\o_2 b, 2\r Q-(\b_1-\o_2 b+\a_{24} b))= ({\pi \mu\over \g(-b^2)})^2{ \g((\b_1-\r Q,\a_3)b)\g((\b_1-\r Q,\a_{24})b)\over \g((\b_1,\a_3)b)\g(Qb+(\b_1-\r Q,\a_{24})b)}\,.
$$
If  we set $(\b_1,\a_3)=0$ -  as in  the case considered in section 2, the  r.h.s. vanishes.

For the analytic continuation of $c(\b_1,\b_2,2\r Q-\b_3)$
one obtains
  \eqna\formTnewga$$\eqalignno{
&C(\b_1,\b_2,2\r Q-\b_3)=
 {
(b^{2e_0 b} \l)^{-(\b_{12}^3, \r)\over b}\zu^3(b) \over \zu((\b_{12}^3, \o_1)\zu((\b_{12}^3,\o_2-\o_1)-Q) }\cr
&
{  \zu((\r Q-\b_1,\a_2)) \zu((\r Q-\b_1,\a_{24}))  \zu((\r Q-\b_2,\a_2)) 
\zu((\r Q-\b_2,\a_{13}))  \over    \zu((\b_{123^*} -2\r Q,\o_2-\o_1) -Q)  \zu((\b_{123^*}-2\r Q,\o_1)) 
 }\cr
 & {1
 \over    \zu((\b_{23}^1,\o_1)-Q) \zu((\b_{23}^1,\o_2-\o_1)) 
} {1 
 \over 
 \zu( (\b_{13^*}^2,\o_2-\o_1)) \zu((\b_{13^*}^2,\o_1)-Q) } & \formTnewga{}\cr
& {\prod_{\a>0} \zu((\b_3-\r Q, \a ))    
\over \zu((\b_{12}^3,\o_3-\o_1)+(\b_3-\r Q, \a_{24})) 
\zu((\b_{123^*},\o_3-\o_1)+(\r Q-\b^*_{3},\a_{24}))  }\cr
&{\zu((\r Q\!-\!\b_1, \a_{3}))  \over \zu((\b_{123^*}\!-\!2\r Q ,\a_2\!-\!\o_2))
 \zu((\b_{12}^3,\a_2\!-\!\o_2))
\zu((\r Q\!-\!\b_1, \a_{3}) + (\b_{12}^3,\o_3\!-\!\o_1))
 \zu((\b_{12}^3,\o_3\!-\!\o_1))}\,
}$$valid for arbitrary $\b_3$ and  $(\b_2, \a_1)=0=(\b_2, \a_3)\,, (\b_1, \a_1)=0. $
Taking the  residue at $(\b_{12}^3, \o_3-\o_1)=0$ and then setting $(\b_1,\a_3)=0$ one  reproduces
the OPE constant $C(\b_1,\b_2,2\r Q-\b_3)$ with $\b_3=\b_3^*$ of  section 2.

Similarly one  derives the analog of the constant \formTnewga{}  with $(\b_1, \a_3)= 0$ and nonzero components  $(\b_1, \a_i)\ne 0\,, i=1,2$.
\bigskip

\noindent
$\bullet\ $ 
The  duality formula \flid{}\  can be used to show that the 4-point Toda functions 
 of the type discussed in section 3 admit an alternative  integral   representation.
 The derivation is  a  certain  generalisation of   the one in \FLa\ for the case of Liouville correlators with one degenerate field, shown to be proportional to a Coulomb Liouville correlator with generic weights; similar consideration appears in \FLc, \FLd. 

Consider  the 4-point function $\la V_{- \o_2 b } (x) V_{\b_1} (0)V_{\b_2}(1)V_{2\r Q-\b_4}(\infty)\ra_{CG}$ 
with vertex highest weights 
$(\b_a,\a_i)=0, i=1,3\,, a=1,2$,  $\b_3=-\o_2 b$  and $(\b_4, \a_i)=-l b\,, i=1,3$  with non-negative integer $l$. We  assume that this is  a 
 Coulomb correlator  with weights satsfying the charge conservation  condition  $-(\b_{123}^4, \o_1)/b=s_1=s_3$,  with a  
 positive integer $s_1$.
 It is  given by the multiple  integral  $I_{s_1, s_2, s_1}(\b_1,\b_2,\b_3)$ 
with   $ s_2-2s_1=-(\b_{123}^4, \o_2-2\o_1)/b=-(\b_4, \a_1)=l$.  
This integral   is converted   recursively with the help of 
\flid{}\ similarly to what   was done above for the 3-point function. Unlike that computation the recursion does not preserve the type of the integral, since at the first step   formula  \flid{}\   is applied in the last integration with respect to $\{t_j^{(3)},j=1,2,\dots, s_3\}$  with $m=2$, i.e., one more double integral is added 
and this structure after the first step is recursively repeated,  yielding for $1\le s \le s_1$   the  integral 
%%
%\eqna\recfor$$\eqalignno{
%&I_{s_1-s, s_2-2s, s_3-s+1; 2}(\b_1^{(s)},\b_2^{(s)}, \, 0)(x)
%=
%\cr
%&\int d\mu_{s_3-s+1}(t^{(3)}) \Phi^{(s)}(x;t^{(3)}) 
%\int d\mu_2(t^{(4)})  D_2(t^{(4)})  \times  &{} \recfor {} 
%\cr
%&
%\prod_{j=1}^{s_3-s+1} \prod_{i=1}^2 |t_j^{(3)} -t_i^{(4)} |^{2b^2} \,
% \prod_{a=1}^2|t_i^{(4)} -x_a|^{2(\b^{(s)}_a-2\r Q,\a_2 )b}|t_i^{(4)} -x|^{-2(\b^{(s)}_3+2\r Q,\a_2 )b }
%}$$
%
\eqna\recfor$$\eqalignno{
&I_{s_1-s, s_2-2s, s_3-s+1; 2}(\b_1^{(s)},\b_2^{(s)}, \, 0; \b_3^{(s)})(x)
: = &{} \recfor {} 
\cr
&\int d\mu_{s_3-s+1}(t^{(3)}) \Phi^{(s)}(x;t^{(3)}) 
\int d\mu_2(t^{(4)})  D_2(t^{(4)}) \prod_{j=1}^{s_3-s+1} \prod_{i=1}^2 |t_j^{(3)} -t_i^{(4)} |^{2b^2}  \times  
\cr
&
%\prod_{j=1}^{s_3-s+1} \prod_{i=1}^2 |t_j^{(3)} -t_i^{(4)} |^{2b^2} \, \prod_{a=1}^2
 |t_i^{(4)}|^{2(\b^{(s)}_1-2\r Q,\a_2 )b} |t_i^{(4)}-1|^{2(\b^{(s)}_2-2\r Q,\a_2 )b} |t_i^{(4)} -x|^{-2(\b^{(s)}_3+2\r Q,\a_2 )b } 
}$$
%Here $x_1=0, x_2=1,x_3=x$,  
where 
\eqn\tildbe{
\b_a^{(s)}= (\b_a +s \o_2 b)\,, a=1,2,3\,   
}
and $\Phi^{(s)}(x;t^{(3)})$ is the integrand of  $I_{s_1-s, s_2-2s, s_3-s+1}(\b_1^{(s)},\b_2^{(s)}, \, 0)(x)$ integrated over the first two sets of variables $\{t_j^{(1)}|, j=1,2 \dots s_1-s\}\,, \{t_i^{(2)}|, i=1,2 \dots s_2-2s\}$ so that 
$$\int   d\mu_{s_3-s+1}(t^{(3)}) \Phi^{(s)}(x;t^{(3)})  =
 I_{s_1-s, s_2-2s, s_3-s+1}(\b_1^{(s)},\b_2^{(s)}, \, 0)(x)\,.$$

Setting 
$s=s_1=s_3$  we obtain the  integral  $I_{0,l,1; \,2}$  up to a constant $\Omega_{s_1, l}(\{\b_a\})$.
In this integral $s_1$ is a parameter appearing in the new weights $\b_a^{(s_1)}$, so the integral can be continued to generic values of $\b_a$. 
The  constant $\Omega_{s_1, l}(\{\b_a\})$ is analytically continued then for  non-integer $s_1
=-(\b_{123}^4, \o_1)/b$ 
keeping $l$ non-negative integer. 
This determines the initial correlator with the  charge conservation  condition dropped
\eqna\equivrepr$$\eqalignno{
&\la V_{- \o_2 b } (x) V_{\b_1} (0)V_{\b_2}(1)V_{2\r Q-\b_4}(\infty)\ra =
\Omega_l(\{\b_a\})   
 |x|^{2b(4Q\r -(\b_1,\a_2))}  |x- 1|^{2b(4Q-(\b_2,\a_2))}\,  \times
   \cr
   &    \qquad \qquad I_{0,l,1; \,2}(\b_1-(\b_{123}^4, \o_1) \o_2, \b_2-(\b_{123}^4, \o_1) \o_2,0;\b_3-(\b_{123}^4, \o_1) \o_2)(x)\,, 
  & {} \equivrepr {}
  }$$
\eqna\cof$$\eqalignno{
&\Omega_l(\{\b_a\})  =  {b^{4 (\b_{123}^{4} ,\o_1)b}b^{2Qbl} \g(-b^2)^{l+1}\over   \g(- Q b+ (\b_{123}^{4} ,\o_1)b )\g( (\b_{123}^{4} ,\o_1)b)}  {\zu(2Q)\over \zu(Q)}\, \times
\cr
&{ \(b^{2(1-b^2)}\l_T\)^{- {(\b_{123}^{4} ,\r)\over b}}\,
 \zu^2(b) \over \zu((\b_{123}^4, \o_1))  \zu((\b_{123}^4,\o_2- \o_1)-Q)}
\prod_{k=0}^{l-1}b^{b(3Q-lb)}\ \g(kb-Qb)\, \times
{} &  \cof{}   \cr
&\prod_{a=1}^2 {\zu((\b_a,\a_2)  \zu((\b_a-\r Q,\a_2)) 
 \over \zu(Q  +(\b_{123}^4-2\b_a, \o_1))   \zu(2Q +(\b_{123}^4-2\b_a, \o_1)) }\, \times
 \cr
&
{ \zu((\b_4-\r Q, \a_{24})) \zu((\b_4-\r Q, \a_{14}))
\over  \zu((\b_{1234}, \o_2-\o_1)-2Q) \zu((\b_{1234}-2\r Q,\o_1)) }\,.
}$$

\noindent
   
The  constant $\Omega_{s_1, l}(\{\b_a\})$  is  recovered as the coefficient of an order two  pole of   
$\Omega_l(\{\b_a\})$ in \cof{}\ at $(\b_{123}^4, \o_1)=-s_1 b\,, s_1\in \CZ_{>0}$. 
The appearance of the pole of order two is due to the fact that $l
$
  is  a non-negative integer. Alternatively the expression in \cof{}\  can be further 
  extended for  generic  values of $l$  so that  the second order pole splits into two first order poles  - 
  then the initial Coulomb representation of the l.h.s. 
is recovered  by a double residue as in \resdoubl{}.
  
  For $l=0$ the integral in  the r.h.s can be interpreted, after integrating over $t^{(3)}$,  as a 
  Liouville Coulomb  integral $I^{(\tilde{b})}_2(\tb_1, \tb_2,\tb_3)(x)$ with a modified parameter  $b^2\!\to\ \tilde{b}^2\!=\!-Qb$, 
  \eqna\tilb$$\eqalignno{  
 & 2\tb_a \tilde{b}=-\b^{(s)}b+2Qb= -(\b_a,\a_2)b +(\b_{123}^4, \o_1)+2Qb\,, a=1,2\,, \cr
 &2\tb_3 \tilde{b}=\b_3^{(s)} b+2Qb=(\b_3,\a_2)b - (\b_{123}^4, \o_1)+2Qb \,, & {} \tilb {} \cr
 &2\tb_4 \tilde{b}= 2(\tb_{123} + 2 \tilde{b}) \tilde{b} =  -(\b_4,\a_2)b -(\b_{123}^4, \o_1)b 
 + 2\,.
 }$$
 
 This integral represents a  4-point function which admits three fusion channels - in agreement with the truncated to three terms Toda fusion rule.  According to the result in \FLa\
 this Coulomb Liouville  4-point function is furthermore related to a Liouville 4-point function  with one degenerate field $V_{-\tilde{b}}$,     which satisfies a third order BPZ differential equation. 
 The observed  relation  between   4-point  functions in the $W_4$ and the 
$W_2$   theory  (with modified parameter $\tilde{b}=-Qb$) suggests that there will be also a relation for the fusing matrices. Indeed, the $3 \times 3$ $F$ matrix transforming the Virasoro  block  with the degenerate vertex highest  weight $\b^L=-\tilde{b}$ and three arbitrary 
representations   has similar  structure to the  $F$ matrix computed in section 3;    the  precise identification will be presented  elsewhere.

  The derivation of the  above  Coulomb representation 
 can be extended to a more 
 general set of weights, e.g.,  restricting only the components   $(\b_a, \a_1)=0\,, a=1,2$
 and  with non-symmetric $\b_4$ s.t.  
 $-(\b_4, \a_1)/b=l \in \CZ_{\ge 0}\, $.
 To ensure that $s_3-s_1 \in \CZ_{\ge 0}$ one has to    impose  additional restrictions on the combination of components $(\b_{12}^4/b,\a_1-\a_3)$.The set includes the doubly  degenerate weights with   $l^{(3)}_a=-(\b_a, \a_3)/b \in \CZ_{\ge 0}\,, a=1,2,3$.

 \listrefs
\bye